\documentclass[twocolumn,tighten,twocolappendix]{aastex701}
\usepackage{graphicx}
\usepackage{amsmath}
\usepackage{amsfonts}
\usepackage{amssymb}
\usepackage{xcolor}

\begin{document}

\title{Supergranulation and poleward migration of the magnetic field at high latitudes of the Sun} 

\author[0000-0002-9270-6785]{L. P. Chitta}
\affiliation{Max-Planck-Institut f\"ur Sonnensystemforschung, Justus-von-Liebig-Weg 3, 37077 G\"ottingen, Germany}
\email[show]{chitta@mps.mpg.de}  

\author[0000-0003-2755-5295]{D. Calchetti}
\affiliation{Max-Planck-Institut f\"ur Sonnensystemforschung, Justus-von-Liebig-Weg 3, 37077 G\"ottingen, Germany}
\email{calchetti@mps.mpg.de}

\author{J. Hirzberger}
\affiliation{Max-Planck-Institut f\"ur Sonnensystemforschung, Justus-von-Liebig-Weg 3, 37077 G\"ottingen, Germany}
\email{hirzberger@mps.mpg.de}

\author[0000-0001-7809-0067]{G. Valori}
\affiliation{Max-Planck-Institut f\"ur Sonnensystemforschung, Justus-von-Liebig-Weg 3, 37077 G\"ottingen, Germany}
\email{valori@mps.mpg.de}

\author[0000-0003-3621-6690]{E. R. Priest}
\affiliation{School of Mathematics and Statistics, University of St Andrews, St Andrews, KY16 9SS, UK}
\email{eric.r.priest@gmail.com}

\author[0000-0002-3418-8449]{S. K. Solanki}
\affiliation{Max-Planck-Institut f\"ur Sonnensystemforschung, Justus-von-Liebig-Weg 3, 37077 G\"ottingen, Germany}
\email{solanki@mps.mpg.de}

\author[0000-0003-4052-9462]{D. Berghmans}
\affiliation{Solar-Terrestrial Centre of Excellence -- SIDC, Royal Observatory of Belgium, Ringlaan -3- Av. Circulaire, 1180 Brussels, Belgium}
\email{david.berghmans@oma.be}

\author[0000-0002-5022-4534]{C. Verbeeck}
\affiliation{Solar-Terrestrial Centre of Excellence -- SIDC, Royal Observatory of Belgium, Ringlaan -3- Av. Circulaire, 1180 Brussels, Belgium}
\email{cis.verbeeck@oma.be}

\author[]{E.~Kraaikamp}
\affiliation{Solar-Terrestrial Centre of Excellence -- SIDC, Royal Observatory of Belgium, Ringlaan -3- Av. Circulaire, 1180 Brussels, Belgium}
\email{emil.kraaikamp@oma.be}

\author[0000-0002-3776-9548]{K. Albert}
\affiliation{Max-Planck-Institut f\"ur Sonnensystemforschung, Justus-von-Liebig-Weg 3, 37077 G\"ottingen, Germany}
\email{albert@mps.mpg.de}

\author[0000-0002-1790-1951]{T. Appourchaux}
\affiliation{Univ. Paris-Saclay, Institut d’Astrophysique Spatiale, UMR 8617, CNRS, B{\^a}timent 121, 91405 Orsay Cedex, France}
\email{Thierry.Appourchaux@universite-paris-saclay.fr}

\author[0000-0002-7318-3536]{F. J. Bail\'en}
\affiliation{Instituto de Astrofísica de Andalucía (IAA-CSIC), Apartado de Correos 3004, E-18080 Granada, Spain}
\affiliation{Spanish Space Solar Physics Consortium (S$^{3}$PC), Spain}
\email{fbailen@iaa.es}

\author[0000-0001-8669-8857]{L. R. Bellot~Rubio}
\affiliation{Instituto de Astrofísica de Andalucía (IAA-CSIC), Apartado de Correos 3004, E-18080 Granada, Spain}
\affiliation{Spanish Space Solar Physics Consortium (S$^{3}$PC), Spain}
\email{lbellot@iaa.es}

\author[0000-0002-2055-441X]{J. Blanco Rodr\'\i guez}
\affiliation{Universitat de Val\`encia, Catedr\'atico Jos\'e Beltr\'an 2, E-46980 Paterna-Valencia, Spain}
\affiliation{Spanish Space Solar Physics Consortium (S$^{3}$PC), Spain}
\email{julian.blanco@uv.es}

\author[]{A. Feller}
\affiliation{Max-Planck-Institut f\"ur Sonnensystemforschung, Justus-von-Liebig-Weg 3, 37077 G\"ottingen, Germany}
\email{feller@mps.mpg.de}

\author[0000-0002-9972-9840]{A. Gandorfer}
\affiliation{Max-Planck-Institut f\"ur Sonnensystemforschung, Justus-von-Liebig-Weg 3, 37077 G\"ottingen, Germany}
\email{gandorfer@mps.mpg.de}

\author[0000-0001-7696-8665]{L. Gizon}
\affiliation{Max-Planck-Institut f\"ur Sonnensystemforschung, Justus-von-Liebig-Weg 3, 37077 G\"ottingen, Germany}
\affiliation{Institut f\"ur Astrophysik, Georg-August-Universit\"at G\"ottingen, Friedrich-Hund-Platz 1, 37077 G\"ottingen, Germany}
\email{gizon@mps.mpg.de}

\author[0000-0003-1459-7074]{A. Lagg}
\affiliation{Max-Planck-Institut f\"ur Sonnensystemforschung, Justus-von-Liebig-Weg 3, 37077 G\"ottingen, Germany}
\email{lagg@mps.mpg.de}

\author[0000-0002-7336-0926]{A. Moreno Vacas}
\affiliation{Instituto de Astrofísica de Andalucía (IAA-CSIC), Apartado de Correos 3004, E-18080 Granada, Spain}
\affiliation{Spanish Space Solar Physics Consortium (S$^{3}$PC), Spain}
\email{amoreno@iaa.es}

\author[0000-0001-8829-1938]{D. Orozco~Su\'arez}
\affiliation{Instituto de Astrofísica de Andalucía (IAA-CSIC), Apartado de Correos 3004, E-18080 Granada, Spain}
\affiliation{Spanish Space Solar Physics Consortium (S$^{3}$PC), Spain}
\email{orozco@iaa.es}

\author[0000-0002-2391-6156]{J. Schou}
\affiliation{Max-Planck-Institut f\"ur Sonnensystemforschung, Justus-von-Liebig-Weg 3, 37077 G\"ottingen, Germany}
\email{schou@mps.mpg.de}

\author[0000-0001-6060-9078]{U. Sch\"uhle}
\affiliation{Max-Planck-Institut f\"ur Sonnensystemforschung, Justus-von-Liebig-Weg 3, 37077 G\"ottingen, Germany}
\email{schuehle@mps.mpg.de}

\author[0000-0002-5387-636X]{J. Sinjan}
\affiliation{Max-Planck-Institut f\"ur Sonnensystemforschung, Justus-von-Liebig-Weg 3, 37077 G\"ottingen, Germany}
\email{sinjan@mps.mpg.de}

\author[0000-0003-1483-4535]{H. Strecker}
\affiliation{Instituto de Astrofísica de Andalucía (IAA-CSIC), Apartado de Correos 3004, E-18080 Granada, Spain}
\affiliation{Spanish Space Solar Physics Consortium (S$^{3}$PC), Spain}
\email{streckerh@iaa.es}

\author[]{R. Volkmer}
\affiliation{Institut f\"{u}r Sonnenphysik (KIS), Georges-K\"{o}hler-Allee 401a, 79110 Freiburg, Germany}
\email{volkmer@leibniz-kis.de}

\author[0000-0001-5833-3738]{J. Woch}
\affiliation{Max-Planck-Institut f\"ur Sonnensystemforschung, Justus-von-Liebig-Weg 3, 37077 G\"ottingen, Germany}
\email{woch@mps.mpg.de}

\author[0000-0001-8164-5633]{X. Li}
\affiliation{Max-Planck-Institut f\"ur Sonnensystemforschung, Justus-von-Liebig-Weg 3, 37077 G\"ottingen, Germany}
\email{lixiaohong@mps.mpg.de}

\author[0000-0002-7044-6281]{T. Oba}
\affiliation{Max-Planck-Institut f\"ur Sonnensystemforschung, Justus-von-Liebig-Weg 3, 37077 G\"ottingen, Germany}
\email{oba@mps.mpg.de}

\author[]{A. Ulyanov}
\affiliation{Max-Planck-Institut f\"ur Sonnensystemforschung, Justus-von-Liebig-Weg 3, 37077 G\"ottingen, Germany}
\email{ulyanov@mps.mpg.de}



\begin{abstract}
Magnetoconvection at the solar surface governs the dynamics in the upper solar atmosphere and sustains the heliosphere. Properties of this fundamental process are poorly described near the solar poles. Here we report the first out-of-ecliptic remote-sensing observations of the south pole of the Sun from a high-latitude campaign of the Solar Orbiter spacecraft which reveal spatial and temporal evolution of supergranular convective cells. The supergranular cells have spatial scales of 20--40\,Mm. From eight days of observations starting on 2025 March 16, our analysis shows that the magnetic network migrates poleward, on average, at high latitudes (above 60\textdegree), with speeds in the range of 10--20\,m\,s$^{-1}$, depending on the structures being tracked. These results shed light on the buildup of the polar magnetic field that is central to our understanding of the solar cycle and the heliospheric magnetic field.

\end{abstract}


\section{Introduction}
Magnetoconvective processes including random buffeting of photospheric magnetic field by convective motions, emergence and cancellation of magnetic flux, play a vital role in building the solar atmosphere \citep[][]{1998Natur.394..152S,2001SoPh..200...23P,2006SoPh..234...41K,2023ApJ...956L...1C}. Solar convection exhibits a granular pattern on scales of $\sim$1\,Mm, with individual features being characterized by brighter upwelling plasma, surrounded by downflowing material at their outer edges in the darker intergranular lanes \citep[][]{1986SoPh..107...11R}. The lifetimes of granular structures is within a range between 1 and 30\,min with an average of 6\,min  \citep[][]{1999ApJ...515..441H}. 

On scales larger than 10\,Mm distinct convective cells with typical lifetimes of 1--2\,day and diameters of 25--30\,Mm are observed \citep[][]{Duvall2000,2004ApJ...616.1242D,2008SoPh..251..417H,2014A&A...567A.138R,2018LRSP...15....6R}, with estimations that depend on the measurement technique \citep[][]{2014masu.book.....P}. These supergranules, characterized by their predominantly horizontal (lateral) outflows with speeds in the range of 300--500\,m\,s$^{-1}$ \citep[][]{1964ApJ...140.1120S,2018LRSP...15....6R}, systematically advect the intergranular magnetic field from cell interior to its boundary, forming the magnetic network \citep[][]{2012ApJ...758L..38O}. Supergranules are slightly brighter at their centers, implying a temperature excess of about 1\,K \citep{Langfellner2016}, consistent with a convective interpretation. However, it remains unclear why supergranulation is the dominant scale of solar convection \citep[][]{2018LRSP...15....6R}, and why its pattern evolution does not exactly track the local flow of the plasma \citep{2003Natur.421...43G}.

\begin{figure*}
 \begin{center}
   \includegraphics[width=\textwidth]{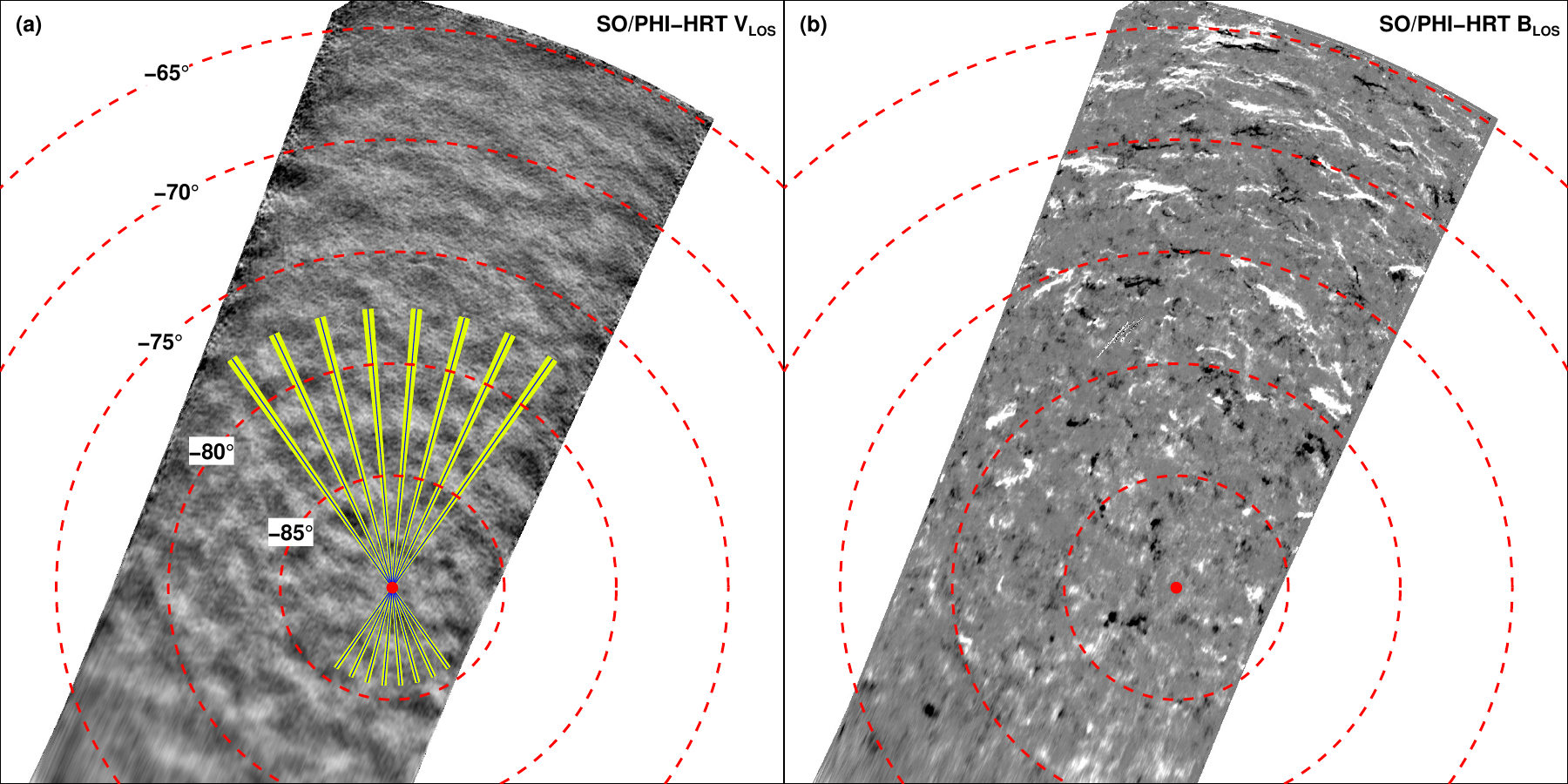}
   \caption{South pole of the Sun. 15-hour temporally averaged maps of the line-of-sight components of the photospheric velocity (a), and magnetic field (b) observed by the SO/PHI-HRT instrument are shown. The maps are displayed in the zenithal equidistant projection, with the zenith placed at the south pole at $-90$\textdegree\ latitude (red dot). The velocity map is saturated at $\pm1$\,km\,s$^{-1}$, with darker (lighter) shaded pixels indicating plasma flows toward (away from) the observer (gray is zero). The magnetic field map is saturated at $\pm5$\,G. Lighter (darker) patches have magnetic field component toward (away from) the observer. In panel (a) the blue colored lines passing through the pole are longitudes, going from 55\textdegree\ to 125\textdegree\ in a counter-clockwise direction, with 10\textdegree\ spacing. The yellow shaded region is a $\pm$1\textdegree\ longitudinal band. The concentric circles in both panels represent lines of latitude with a separation of 5\textdegree. An animation of panel (b) at 2-hour cadence is available online. The animation has a play back time of 2\,s\ and contains 7 frames with time stamps from 2025\,March\,21\,UT\,10:29 to 2025\,March\,21\,UT\,22:29, with $\sim$2\,hour increments. See Appendix\,\ref{app:phi} for more details.  \label{fig:phi_map}}
 \end{center}
\end{figure*}

Beyond resolving their origin, deciphering the nature of supergranulation at high latitudes is also particularly crucial to our understanding of the transport and long-term build-up of the magnetic field in the polar regions through meridional circulation. On the one hand, the strength of the polar magnetic field is a good predictor of the amplitude of the following cycle \citep[][]{1978GeoRL...5..411S,2013ApJ...767L..25M}. On the other, the polar field is also important for the strength of the heliospheric magnetic field, much of which is rooted in polar coronal holes at solar minimum \citep[][]{1995Sci...268.1007B}. Additionally, network magnetic field evolution governed by supergranular flows in open-field coronal hole regions can release the plasma into the solar wind, inflating the heliosphere \citep[][]{1999Sci...283..810H,2014Sci...346A.315T,2023Sci...381..867C}. However, the structure and evolution of supergranulation at high latitudes near polar regions of the Sun has been elusive due to lack of direct observations. Here we report the first out-of-ecliptic measurements of supergranular-scale magnetoconvection, particularly the size, timescale, and its latitudinal migration at high latitudes near the south pole of the Sun using unprecedented observations from the Solar Orbiter Spacecraft \citep[see, e.g.,][]{2020A&A...642A...1M}.

\begin{figure*}
 \begin{center}
   \includegraphics[width=\textwidth]{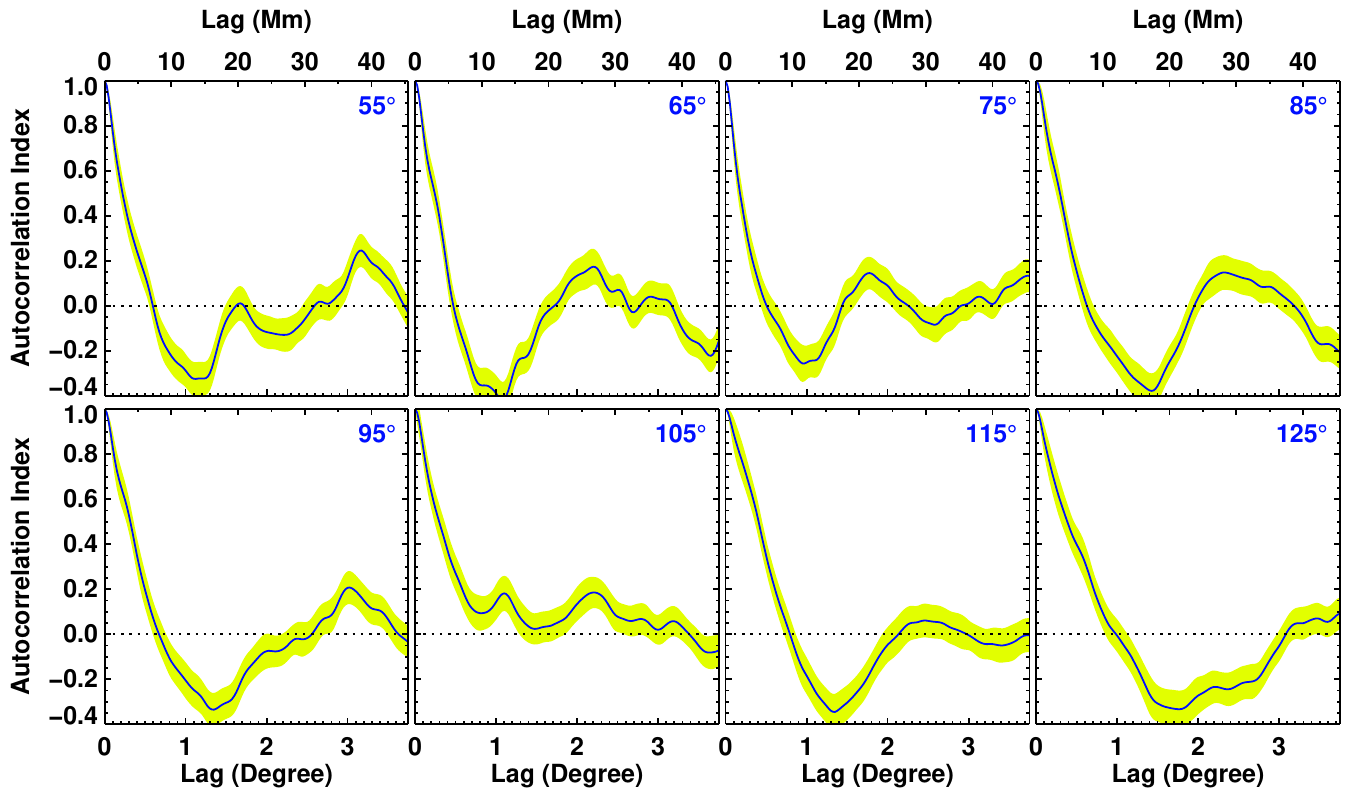}
   \caption{Spatial scales of supergranules near the poles. In each panel we plot the spatial autocorrelation function of the mean line-of-sight velocity (blue colored curves) from the corresponding meridian band in Fig.\,\ref{fig:phi_map}, as labeled. This autocorrelation function is plotted as a function of spatial lag in degrees (lower abscissa) and megameters (upper abscissa). The yellow shaded curves are the 1$\sigma$ standard deviation in the autocorrelation signal. The dotted horizontal lines mark the zero of the autocorrelation function. See Appendix\,\ref{app:auto} for more details.\label{fig:acorr}}
 \end{center}
\end{figure*}

\begin{figure*}
 \begin{center}
   \includegraphics[width=\textwidth]{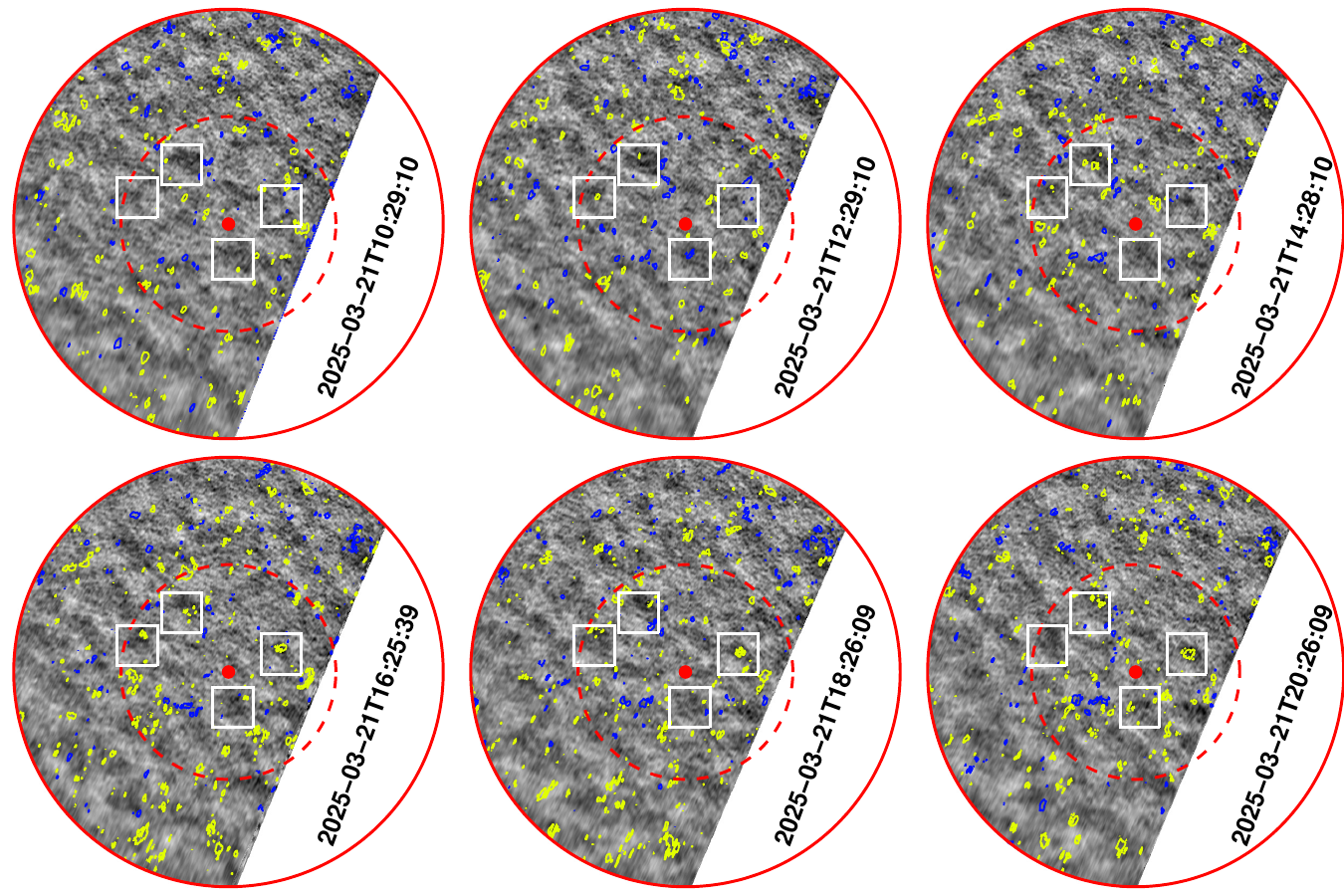}
   \caption{Supergranular structure and evolution near the pole. An image sequence showing the 2-hour time-averaged maps of the line-of-sight components of the photospheric velocity, in the same projection as in Figure\,\ref{fig:phi_map}a. The maps are saturated at $\pm1.5$\,km\,s$^{-1}$. The dashed and solid circles are $-85$\textdegree\ and $-80$\textdegree\ latitudes, respectively. The yellow and blue contours outline the line-of-sight component of the magnetic field (above 7\,G) directed toward and away from the observer. The white squares outline a sample of four distinct supergranular cells near the pole. An animation of this figure is available online. The animation has a play back time of 2\,s\ and contains 7 frames with time stamps from 2025\,March\,21\,UT\,10:29 to 2025\,March\,21\,UT\,22:29, with $\sim$2\,hour increments. \label{fig:super}}
 \end{center}
\end{figure*}

\section{Observations}
\label{sec:obs}
The remote-sensing data of the south pole analyzed in this study were obtained as a part of a pair of Solar Orbiter Observing Plans \citep[SOOP;][]{2020A&A...642A...3Z}, when the spacecraft was at a heliographic latitude of $-16.8$\textdegree. In particular, we used spectropolarimetric observations of the photospheric Fe\,{\sc i} 6173\,\AA\ line recorded by the High Resolution Telescope (HRT) of the Polarimetric and Helioseismic Imager  \citep[SO/PHI;][]{2020A&A...642A..11S} to probe the line-of-sight components of the surface plasma flow field and magnetic field. We combined these with the extreme ultraviolet (EUV) images from the 304\,\AA\ filter of the full-Sun imager (FSI304) on the Extreme Ultraviolet Imager \citep[EUI;][]{2020A&A...642A...8R}, to analyze the chromospheric and transition region imprints of the magnetic network. 

During the observing campaigns the SO/PHI-HRT instrument recorded the south pole of the Sun over a duration of 15\,hours, starting on 2025\,March\,21 at 09:30 universal time (UT). The EUI/FSI304 images considered in this study span 8\,days of observations starting on 2025\,March\,16. Additional details of these observations and data processing methods are provided in Appendices\,\ref{app:phi} and \ref{app:eui}.

\section{Polar supergranular magnetoconvection}
\label{sec:struct}

We begin with the investigation of the spatial structuring of supergranular-scale convection and the magnetic network at the poles. To visualize these structures, we present temporal averages of the SO/PHI-HRT line-of-sight velocity and magnetic field data over the full 15\,hour period, both mapped to the zenithal equidistant projection (Fig.\,\ref{fig:phi_map}). Owing to their horizontal flows from the cell interiors, the supergranules can be seen as corrugated or ripple-like darker and lighter shaded regions in the velocity map, indicating flows toward and away from the observer (Figs.\,\ref{fig:phi_map}a and \ref{fig:vlos}). At latitudes of $-80$\textdegree\ and lower, the magnetic network is seen as latitudinal arcs tracing the solar rotation (Fig.\,\ref{fig:phi_map}b). These arcs can also be seen with chromospheric signatures of the magnetic network over the course of 8\,days of EUI observations (Fig.\,\ref{fig:euiavg}).

To determine the structure and scale of supergranulation near the pole we computed the spatial autocorrelation of the line-of-sight velocity signal. For a better estimate of these sizes and the associated errors, we carried out the autocorrelation analysis at 8 distinct meridian bands (yellow shaded regions in Fig.\,\ref{fig:phi_map}; details given in Appendix\,\ref{app:auto}). The results are displayed in Fig.\,\ref{fig:acorr}. In most cases, the autocorrelation signal falls off steeply to negative values in the lag range of 10--20\,Mm first, before turning to positive values. This turnover sets a minimum spatial scale of the coherent velocity signal that we detect. The autocorrelation functions show distinct positive peaks at various spatial scales in the 20\,Mm to 40\,Mm range. The polar supergranules thus seem to show sizes comparable to the ones at lower latitudes where the  mean diameter is 25\,Mm \citep[][]{2014A&A...567A.138R}.

As shown in Fig.\,\ref{fig:phi_map}, temporal averaging of the data will better reveal the supergranulation. We found that these structures become discernible already at 2\,hours worth of data averaging. We analyzed the temporal evolution of supergranulation near the pole with a 2-hour cadence (Fig.\,\ref{fig:super}; Movie S2). This image sequence shows the supergranular cells, as observed in the line-of-sight velocity maps, expanding, disappearing and new features emerging on these 2-hour timescales. It will be interesting in the future to obtain the lifetime of supergranulation at the polar regions in comparison to the low latitude structures. Longer time series of SO/PHI-HRT observations of poles and near-limb observations at lower latitudes at similar cadences are needed to quantitatively estimate and compare the latitudinal behavior of supergranulation. 

\section{Poleward migration of the magnetic network}
\label{sec:pole}

\begin{figure*}
 \begin{center}
   \includegraphics[width=\textwidth]{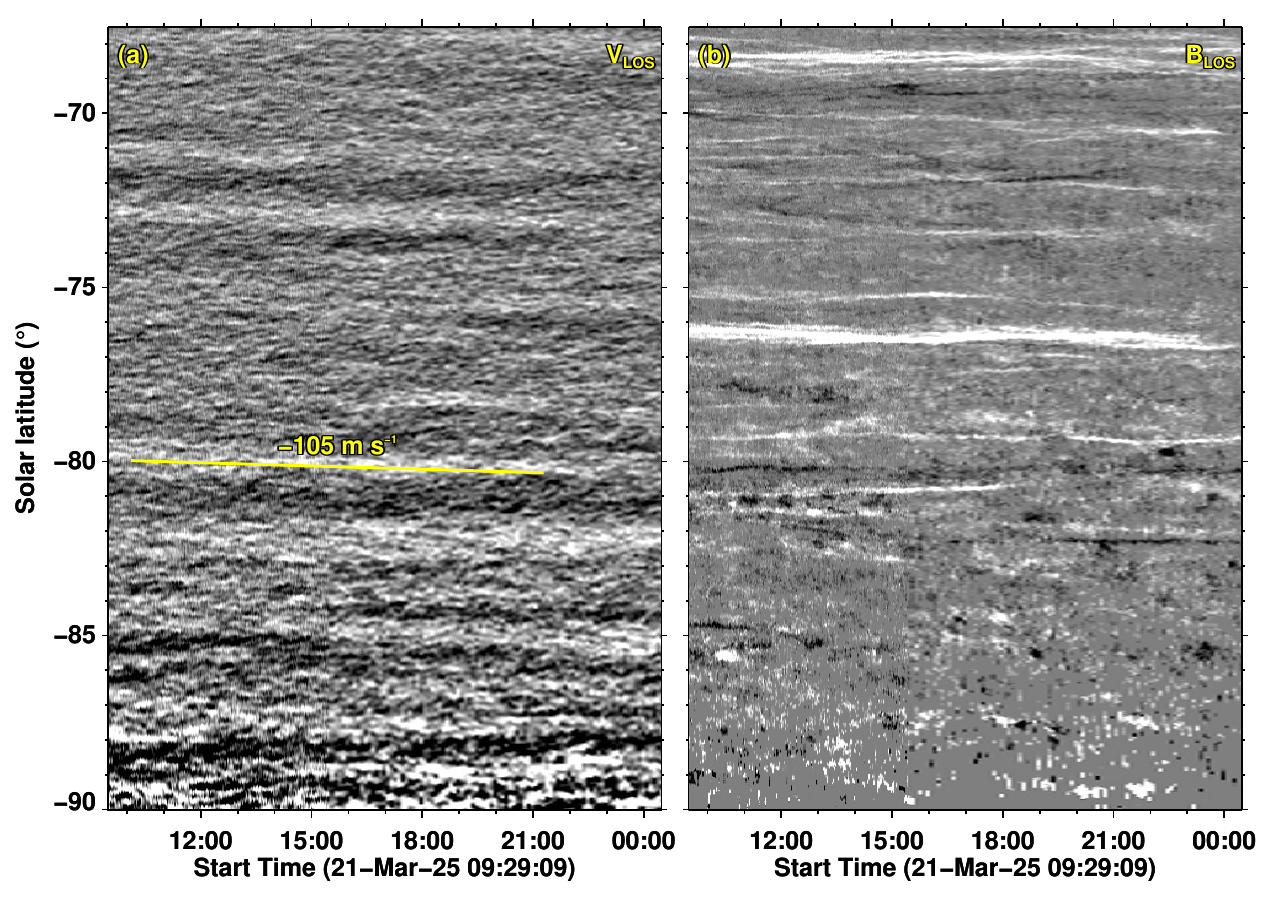}
   \caption{The latitudinal migration of supergranular structures. Latitudinal signal of the longitudinally-averaged line-of-sight components of the velocity (a), and magnetic field (b) are displayed as functions of time. In panel (a) we highlight a feature that is drifting toward the pole with a velocity of $-105$\,m\,s$^{-1}$. The negative sign indicates that the flow is away from the north pole, i.e., toward the south pole. See Appendix\,\ref{app:auto} for more details. \label{fig:phitrc}}
 \end{center}
\end{figure*}

\begin{figure*}
 \begin{center}
   \includegraphics[width=\textwidth]{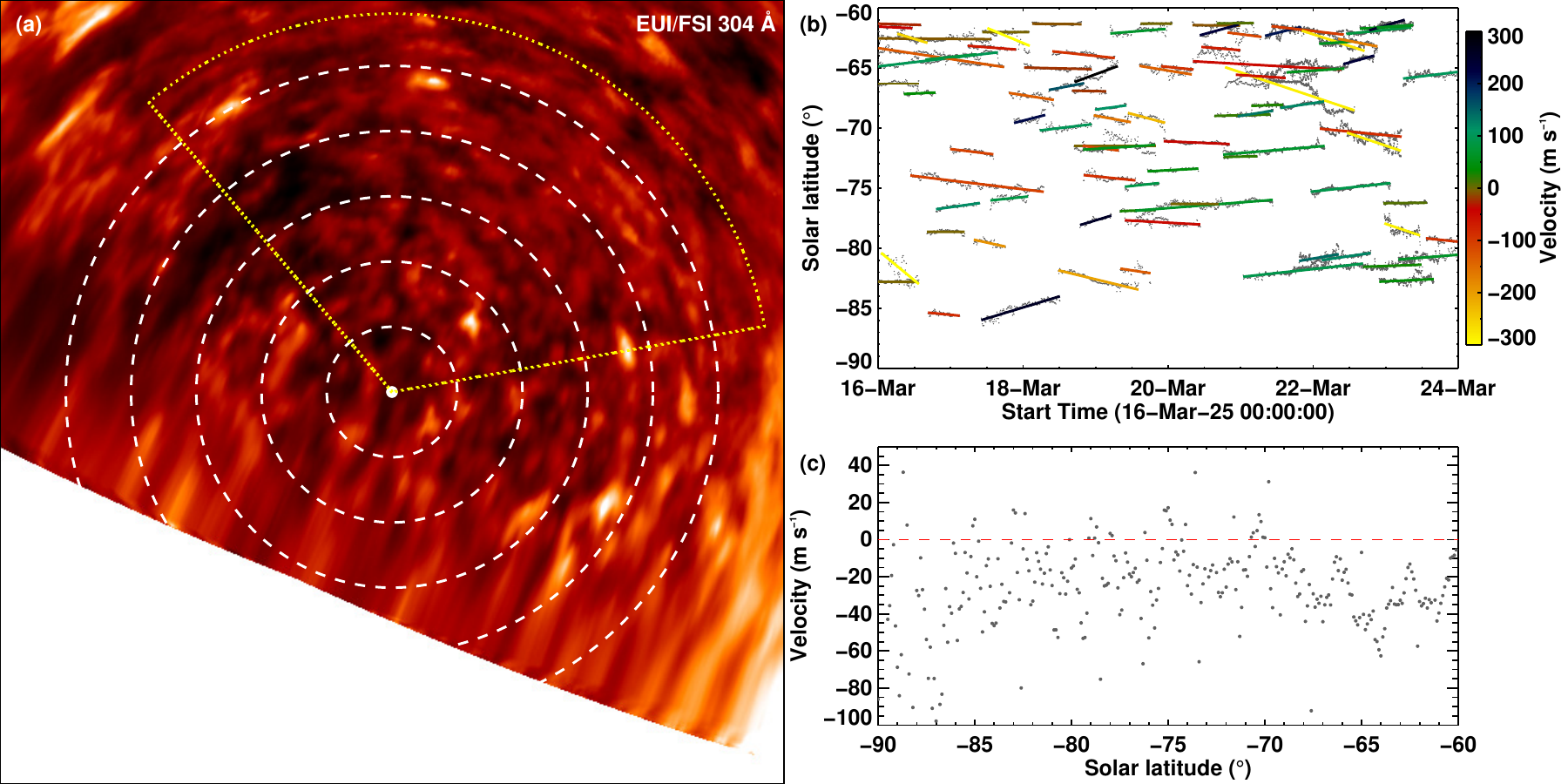}
   \caption{Latitudinal migration of the magnetic network at high latitudes. (a) A 15-hour temporally averaged EUI/FSI 304\,\AA\ intensity map in the zenithal equidistant projection is shown. The south pole is identified by the white dot, with dashed concentric circles denoting latitudes with 5\textdegree\ spacing. The yellow dotted wedge subtends an angle of 120\textdegree\ at the pole. EUV bright points within this wedge are tracked for their motion on the Sun. (b) Latitudinal tracks of EUV bright points with lifetimes of at least 10\,hours (90 in total) are plotted with gray symbols. The various colored solid lines are linear fits to the observed tracks. The colors themselves indicate the latitudinal velocity of the moving bright points (we assign negative values for the bright points moving toward the south pole). (c) Longitudinally and temporally averaged velocity signal, derived from a Fourier local correlation tracking technique, as a function of solar latitude is plotted as gray symbols. The red  horizontal line marks the $0$\,m\,s$^{-1}$ level. In this representation too, negative velocities correspond to motions toward the south pole. See Appendices\,\ref{app:eui} and \ref{app:bright} for more details.\label{fig:euitrc}}
 \end{center}
\end{figure*}

We probed the latitudinal migration of supergranular cells and the magnetic network. The latitude vs. time maps of the SO/PHI-HRT data (details provided in Appendix\,\ref{app:auto}) reveal tracks of individual features of supergranulation and magnetic network. These features show a tendency for migration toward or away from the pole. We highlight one such supergranular structure in the line-of-sight velocity map showing a poleward migration with speed of 105\,m\,s$^{-1}$ (Fig.\,\ref{fig:phitrc}a). In the magnetic field map tracks both toward and away from the pole are also apparent (Fig.\,\ref{fig:phitrc}b). But given that the SO/PHI instrument observed the pole for only 15\,hours, the systematic, long-term nature of this migratory pattern of the supergranulation and the magnetic network cannot be assessed from these data. 

We complemented the SO/PHI data with EUI/FSI304 images from the 8-day observing sequence (Fig.\,\ref{fig:euitrc}a; details given in Appendix\,\ref{app:eui}). The magnetic network forms bright points in the chromosphere. Because these bright points are typically advected by supergranular flows \citep[][]{2009A&A...495..319I}, their long-lived nature (longer than typical granular average lifetimes of 6\,minutes) will enable tracing the horizontal flows over long periods of several hours. Here we tracked 90 such prominent chromospheric network bright points during the 8-day period, all of which lived for at least 10\,hours (Appendix\,\ref{app:bright}). The longitudinal tracks of these bright points is shown in Fig.\,\ref{fig:euitrc}b. As is evident from the plot, the motions of these bright points at high latitudes are distributed both toward and away from the pole. The derived speeds are consistent with the supergranular horizontal flows \citep[][]{2018LRSP...15....6R}. The motion of these bright points, as determined by the summation of all the slopes in Fig.\,\ref{fig:euitrc}c, however, is a net poleward migration with a speed of $10$\,m\,s$^{-1}$.

We have further complemented these measurements by tracking the chromospheric network structure as a whole instead of restricting our analysis to localized bright points, using a Fourier local correlation tracking technique (Appendix\,\ref{app:bright}). The longitudinally and temporally averaged network flow as a function of latitude is shown in Fig.\,\ref{fig:euitrc}c. The mean flow is poleward at high latitudes. After excluding the innermost $5$\textdegree\ latitudinal range from the pole where the flow signal is highly fluctuating, our analysis reveals that there is a net poleward flow of the magnetic network of about $20$\,m\,s$^{-1}$, when averaged over the latitude.

\section{Discussion and Conclusion}

There could be multiple reasons for the observed poleward migration of the magnetic network with speed in the range of 10--20\,m\,s$^{-1}$. These possibilities are discussed in the following.

Could it be that the flow is related to the noise arising from the turbulent supergranular motion itself? A typical supergranule has horizontal flow speeds in the range of 300--500\,m\,s$^{-1}$. Based on the area covered by the wedge in Fig.\,\ref{fig:euitrc}a (ignoring curvature effects), and the average size of a supergranule, we estimate that there will be about 300 supergranular cells in that region. Assuming a lifetime of 1\,day, in the course of 8\,days, there will be a total of about 2400 supergranules in the wedge region. Then the standard error in the flow speed measurement, that is the horizontal flow speed divided by $\sqrt{2400}$, amounts to 6--10\,m\,s$^{-1}$. This would then imply that the motion of all the supergranules over the 8-day period must somehow yield noise levels of 6--10\,m\,s$^{-1}$ due to incomplete cancellation of supergranular velocities. 
This unavoidable supergranulation noise was predicted before the launch of Solar Orbiter by \citet[][their figure 3]{Loeptien2015} at the level of $\sim (\cos \lambda)^{-1/2} \times 1$\,m\,s$^{-1}$ for a 30-day time series at latitude $\lambda$, which corresponds to   $7$~m/s for an 8-day time series $5^\circ$ away from the pole. Thus this noise level is  small to exclude the possibility of a large-scale flow coherently transporting supergranules at speeds in the range of 10--20\,m\,s$^{-1}$.  

We note that the supergranulation exhibits oscillations with a period of 6--9\,days and a pattern propagation  that is  prograde near the equator and equatorward at higher latitudes \citep[][]{2003Natur.421...43G,2003ApJ...596L.259S,2018A&A...617A..97L}. The wave-like motion of the pattern at high polar latitudes is however not known, as it would require much longer time series to be measured. 

Another component of motion that should be considered is the one associated with the high-latitude global inertial modes \citep[][]{2021A&A...652L...6G}. In particular, the $m=1$ high-latitude mode has a cross-polar flow component \citep{2025A&A...695A..67L} that could potentially contribute to our observations (with a sign depending on longitude). However this mode has an amplitude below 5\,m\,s$^{-1}$ during periods of maximum solar activity \citep{2025A&A...695A..67L}. This effect could, however, become important in advecting polar magnetic field during the solar minimum \citep[][]{2025arXiv251000596H}.

Finally, there is a large-scale meridional circulation on the Sun, an axisymmetric flow which transports the surface magnetic field from equatorial regions toward the poles \citep[][]{1979SoPh...63....3D,2022LRSP...19....3H}. The poleward meridional circulation at the surface and an equator-ward return flow in the interior of the convection zone are important features in predicting and modeling the length and strength of the solar cycle \citep[][]{2010Sci...327.1350H,2012ApJ...761L..14R,2014ApJ...782...93H,2020Sci...368.1469G}. 

Previous results of the meridional circulation were obtained from the ecliptic plane view of the Sun. These include, for instance, detection of the poleward flow based on tracking small surface magnetic elements \citep[][]{1993SoPh..147..207K,2010Sci...327.1350H,2012ApJ...761L..14R}  and helioseismology \citep[e.g.][]{2020Sci...368.1469G}. These observations typically show that the meridional circulation peaks at mid-latitudes, reaching poleward speeds of about 15\,m\,s$^{-1}$ in both hemispheres and tends toward zero or lower speeds at higher latitudes near poles \citep[][]{1993SoPh..147..207K,2010Sci...327.1350H,2012ApJ...761L..14R}. Some surface flux transport models assume that the meridional flow tends toward zero at latitudes around 75\textdegree\ \citep[][]{1998ApJ...501..866V,2024ApJ...970..183Y}. Ground-based Doppler shift observations of the Fe\,{\sc i}\,5250\,\AA\ line have revealed a reversed meridional circulation pattern near polar regions for three successive solar minima \citep[][]{2010ApJ...725..658U}. The meridional circulation at very high latitudes may vary significantly depending on the phase of the cycle. There have also been suggestions of a counter-cell meridional flow near polar regions with an amplitude of 3\,m\,s$^{-1}$ \citep[][]{2024ApJ...970..183Y}, based on the ecliptic view of the Sun. Here our observations from a high-latitude perspective with a much better view of the polar region reveal a trend of flows that is poleward at all latitudes in contrast to some earlier studies. But how this trend evolves as the cycle progresses from the current phase of activity maximum toward minimum needs to be examined.

The observed poleward flows at high latitudes that we see are a factor of 2 to 4 faster than the meridional circulation based on tracking magnetic features at the solar surface \citep[][]{2012ApJ...761L..14R}. Here we tracked the EUV signatures of the magnetic field in the form of chromospheric network elements. If the flows that we identified are indeed related to the meridional circulation and are faster, then this potentially suggests a radial gradient in the transport of the magnetic field from the surface through the chromosphere. 

How and which of these different effects contribute to the observed poleward migration of the magnetic field requires further scrutiny. But they all hint at an underlying new aspect of the magnetic field transport at high latitudes. The results here (and in Calchetti et al. 2025 in prep) gave the first concrete strategy for the designing of future campaigns such that they can provide the observational support to answer the remaining questions pertaining to the solar poles. In particular, future high-latitude campaigns of the Solar Orbiter will continue to shed light on the long-term nature of the poleward migration of the magnetic field that we have detected here and its dependence on the phase of the solar cycle. Overall, these Solar Orbiter high-latitude campaigns herald a new era in the exploration of the polar regions of the Sun. Our observations also provide new inputs to future polar mission concepts \citep[][]{2023BAAS...55c.160H, 2023BAAS...55c.333R,2025arXiv250620502D}.

\begin{acknowledgements}
The authors thank the anonymous referee for constructive comments. L.P.C. thanks Robert Cameron (MPS) for helpful discussions. This project has received funding from the European Research Council (ERC) under the European Union's Horizon Europe research and innovation programme (grant agreement Nos. 10103984 -- project ORIGIN; 101097844 -- project WINSUN; 810218 -- Synergy project WHOLE SUN). We also acknowledge funding received under the Horizon Europe programme of the European Union (grant agreement no. 101131534 -- project DynaSun). Solar Orbiter is a space mission of international collaboration between ESA and NASA, operated by ESA. We thank the ESA SOC and MOC teams for their support. The EUI instrument was built by CSL, IAS, MPS, MSSL/UCL, PMOD/WRC, ROB, LCF/IO with funding from the Belgian Federal Science Policy Office (BELSPO/PRODEX PEA 4000134088, 4000106864 and 4000112292); the Centre National d’Etudes Spatiales (CNES); the UK Space Agency (UKSA); the Bundesministerium für Wirtschaft und Energie (BMWi) through the Deutsches Zentrum für Luft- und Raumfahrt (DLR); and the Swiss Space Office (SSO). EK and CV thank the Belgian Federal Science Policy Office (BELSPO) for the provision of financial support in the framework of the PRODEX Programme of the European Space Agency (ESA) under contract number 4000134088. The German contribution to SO/PHI is funded by the BMWi through DLR and by MPG central funds. The Spanish contribution is funded by AEI/MCIN/10.13039/501100011033/ and European Union ``NextGenerationEU''/PRTR'' (RTI2018-096886-C5, PID2021-125325OB-C5, PCI2022-135009-2, PCI2022-135029-2) and ERDF ``A way of making Europe''; ``Center of Excellence Severo Ochoa'' awards to IAA-CSIC (SEV-2017-0709, CEX2021-001131-S); and a Ram\'on y Cajal fellowship awarded to DOS. The French contribution is funded by CNES. 
\end{acknowledgements}

\bibliographystyle{aasjournalv7}

\appendix

\section{SO/PHI-HRT data}
\label{app:phi}
The Polarimetric and Helioseismic Imager instrument \citep[SO/PHI;][]{2020A&A...642A..11S} houses the High Resolution Telescope (HRT) that records spectropolarimetric measurements of the photospheric Fe\,{\sc i} 6173\,\AA\ line. The four Stokes parameters ($I$, $Q$, $U$, and $V$) are sampled at five wavelength positions within the Fe\,{\sc i} line, along with a nearby continuum point. More details on the SO/PHI instrumentation and data calibration are given in \citet{2018SPIE10698E..4NG,2022SPIE12180E..3FK,2022SPIE12189E..1JS,2023A&A...675A..61K,2024A&A...681A..58B}. 

As a part of the \texttt{R\_SMALL\_HRES\_HCAD\_RS-burst} SOOP, SO/PHI-HRT observed the south pole from 2025\,March\,21 universal time (UT, measured at the spacecraft)\ 09:30 to UT\,15:28. These photospheric data were obtained at a cadence of 120\,s. With the \texttt{R\_SMALL\_MRES\_MCAD\_Supergranulation} SOOP, SO/PHI-HRT recorded data from 2025\,March\,21 UT\,15:35 to 2025\,March\,22 UT\,00:29, at 360\,s cadence. The HRT instrument has an image scale of 0.5\,arcseconds\,($''$)\,pixel$^{-1}$. During the period of observations, distance of the Solar Orbiter spacecraft from the Sun was about 0.36 astronomical units\ (au). Accordingly, the image scale was about 130\,km\,pixel$^{-1}$ (at disk center). The heliographic latitude was about $-16.8$\textdegree.

We used line-of-sight components of the photospheric magnetic field and velocity information retrieved from the SO/PHI-HRT data. The line-of-sight component of the magnetic field that is directed toward or away from the observer is determined by the Stokes\,$V$ signal. Because the observations covered the south pole when the heliographic latitude was about $-16.8$\textdegree, the radial field has a weak signature in Stokes\,$V$. Consequently,  the line-of-sight component of the photospheric magnetic field is rather weak. We followed the technique described in \citet{2023ApJ...956L...1C} to filter out noise from the magnetic field data. The root-mean-squared fluctuations of the line-of-sight magnetic field have a standard deviation of about 10\,G. Considering this to be the noise level in the data, in each snapshot we first extract all the regions containing at least 9 contiguous pixels per patch, with a signal of at least 3 times this noise level (i.e., more strong line-of-sight field regions). Similarly, we extract regions with signal between 2 and 3 times the noise level (weak field regions), but consider only those patches with at least 25 contiguous pixels. With the imposition of this signal strength and size constraints, we consider that the retained magnetic flux concentrations in our data are reliable for further analysis.

\begin{figure}
 \begin{center}
   \includegraphics[width=0.45\textwidth]{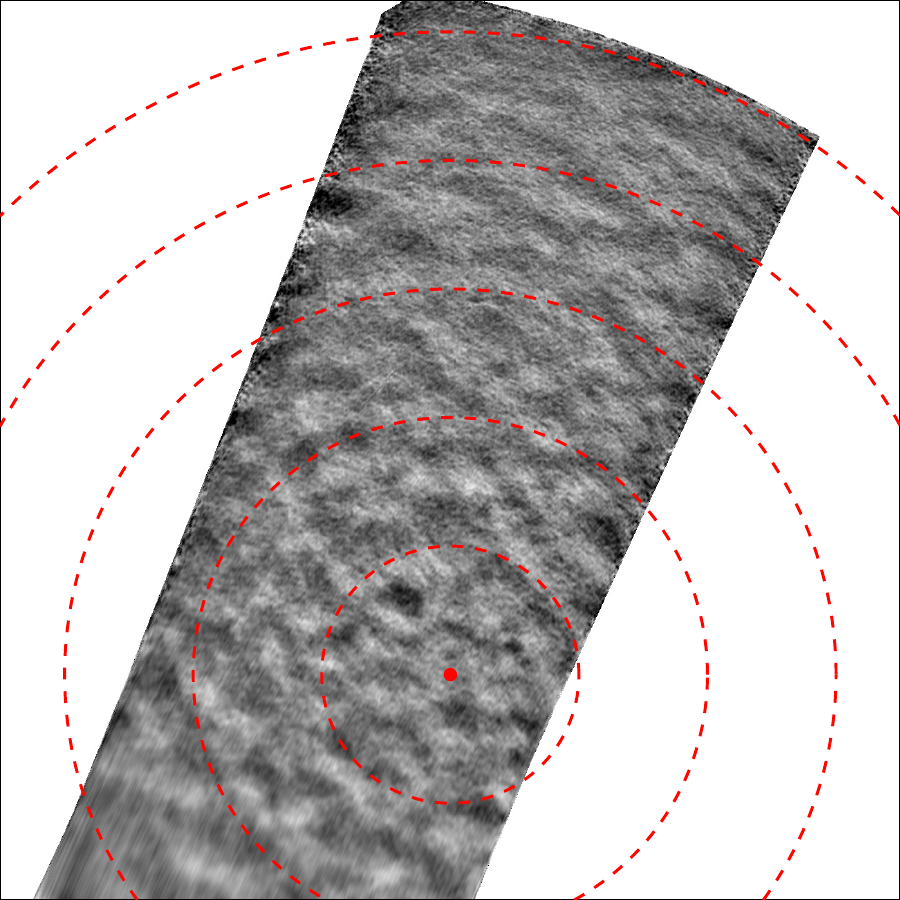}
   \caption{Same as Fig.\,\ref{fig:phi_map}a, but without the lines of longitude overplotted to better visualize the ripple-like velocity structures.\label{fig:vlos}}
 \end{center}
\end{figure}

We remapped these data from the original Helioprojective-Cartesian {\sc tan} projection to the Heliographic zenithal equidistant ({\sc arc}) projection using the World Coordinate System \citep[WCS;][]{2006A&A...449..791T} metadata in our observations. The zenith is placed at the south pole. In this projection, the meridians are straight lines that pass through the zenith. This mapping is most suitable for studying the scales and migrations of various features along the lines of longitude because the distances can be accurately measured along these lines. For this remapping we set the SO/PHI-HRT image scale to be 0.025\textdegree ($\equiv$303\,km at the disk center along the central meridian), which correctly represents the distance when measured from the zenith, i.e., the south pole in this case. Using a wavelet-filtering technique \citep[][]{1994A&A...288..342S}, we removed large-scale gradients (of the size of the field of view) from the {\sc arc} projected line-of-sight velocity maps.

We then temporally averaged spatial maps of the magnetic field and velocity. First, we temporally averaged the whole 15\,hours of the SO/PHI-HRT data for visualization purpose. The corresponding images are displayed in Figs.\,\ref{fig:phi_map} and \ref{fig:vlos}. Second, by temporally averaging data in 2-hour bundles, we also created a time sequence of the SO/PHI-HRT data with 2-hour cadence (shown in Fig.\,\ref{fig:super}), to investigate the evolution of the photospheric flow field and the magnetic field.

\subsection{SO/PHI-HRT data -- Autocorrelation analysis and space-time maps}
\label{app:auto}

We carried out the autocorrelation analysis to determine the sizes of the supergranular cells along eight meridian bands. Each band (any single yellow-shaded region in Fig.\,\ref{fig:phi_map}) comprises 41 meridians spaced at 0.05\textdegree. We derived the mean velocity signal as a function of latitude by first interpolating the line-of-sight velocity from all the snapshots in the 15\,hour long time sequence along all the 41 meridians in a given band, and then computing the temporal and longitudinal averages. The spatial autocorrelation function of this mean velocity signal corresponding to the eight meridian bands are plotted in Fig.\,\ref{fig:acorr}. We estimated the level of variations in these autocorrelation functions in each band by computing the standard deviation ($\sigma$) of all the autocorrelation functions of the original, i.e., unaveraged velocity signals in that respective band. The 1$\sigma$ level in these fluctuation is shown as the yellow-shaded curve in Fig.\,\ref{fig:acorr}. 

We created the SO/PHI-HRT latitude vs. time velocity and magnetic field maps to study the latitudinal migration of photospheric features. The maps shown in Fig.\,\ref{fig:phitrc} are obtained by longitudinally averaging the respective signal over a band of meridians within $\pm$10\textdegree\ about the 75\textdegree\ meridian. The band has a spacing of 0.05\textdegree.

\section{EUI/FSI data analysis}
\label{app:eui}

We used the 304\,\AA\ imaging data from full-Sun imager (FSI304) of the Extreme Ultraviolet Imager instrument \citep[EUI;][]{2020A&A...642A...8R} to investigate the structures at the south pole of the Sun. The passband captures emission due to the He\,{\sc ii} ion, with formation temperature of log$T$\,[K]\,$\approx$\,4.7. The line samples emission from the chromospheric and transition region structures at the base of the solar corona.

EUI/FSI304 carried out the full-disk observations (including the south pole) for a longer period of time and more frequently than the SO/PHI-HRT instrument did. Here we considered eight days of EUI observations starting from 2025\,March\,16. The data up to 2025\,March\,21 UT\,07:50 had 600\,s cadence and after that the cadence increased to 180\,s. The angular image-scale of EUI/FSI304 instrument is 4.44$''$\,pixel$^{-1}$. During this period of eight days, the distance from the Sun of the Solar Orbiter spacecraft decreased from 0.436\,au to 0.338\,au. Accordingly, the image scale changed from 1.4\,Mm to 1.1\,Mm.  In our analysis we made use of the Level-2 data of the EUI Data Release 6 \citep[][]{euidatarelease6}.

\begin{figure}
 \begin{center}
   \includegraphics[width=0.45\textwidth]{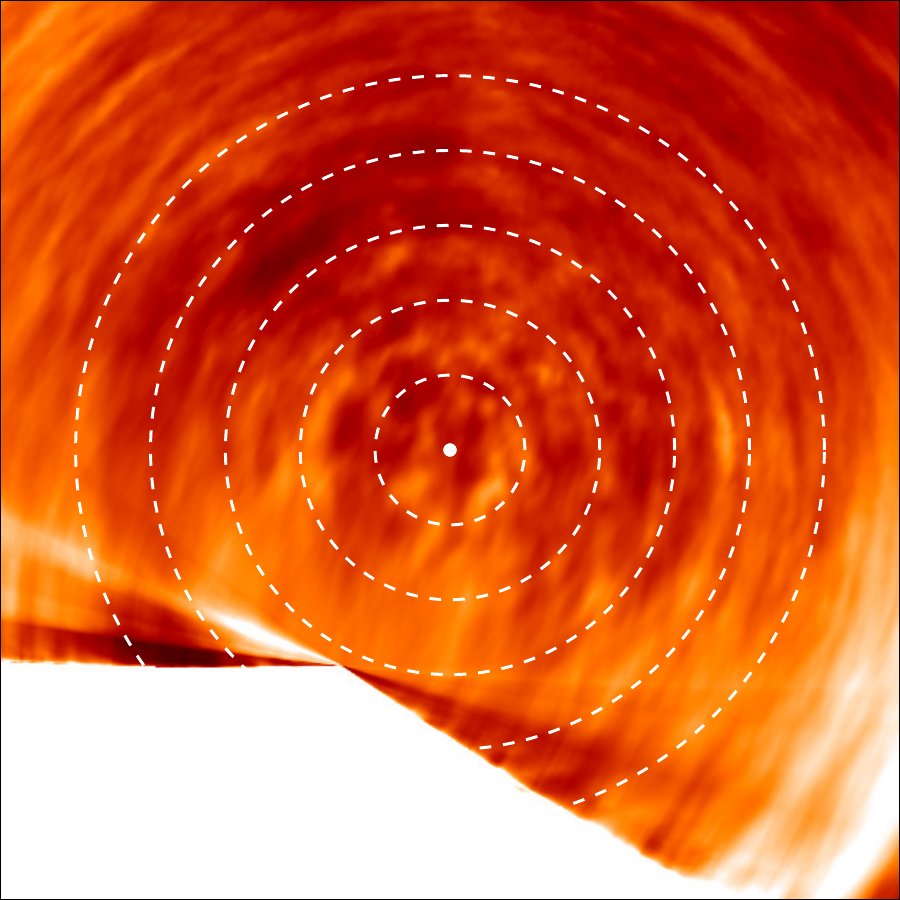}
   \caption{Same as Fig.\,\ref{fig:euitrc}a, but showing the FSI304 map obtained by averaging the whole 8-day dataset.\label{fig:euiavg}}
 \end{center}
\end{figure}

Similar to the SO/PHI-HRT data, we remapped the EUI/FSI304 images to the zenithal equidistant ({\sc arc}) projection, with image scale set to 0.1\textdegree. A 15-hour temporally averaged EUI/FSI304 image corresponding to the period of SO/PHI-HRT observations is shown in Fig.\,\ref{fig:euitrc} and the 8-day average map is plotted in Fig.\,\ref{fig:euiavg}. All averaging is done without compensating for solar rotation.

\subsection{Tracking supergranular flows with the EUI/FSI data}
\label{app:bright}
\begin{figure}
 \begin{center}
   \includegraphics[width=0.45\textwidth]{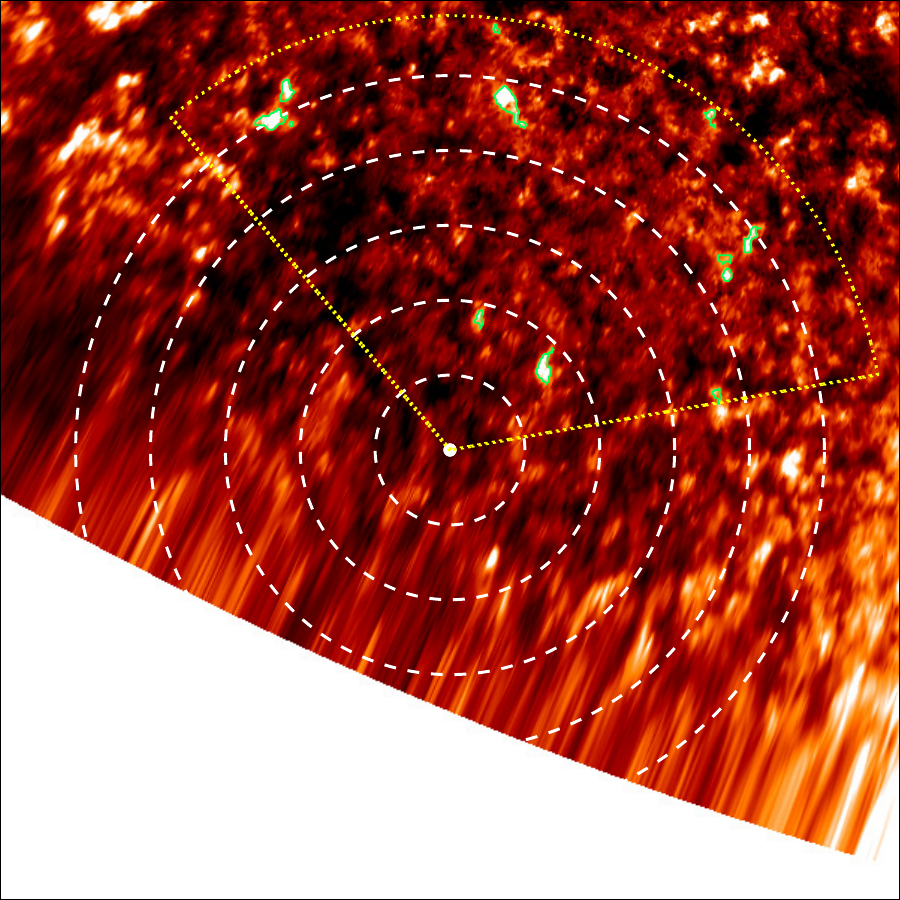}
   \caption{Chromospheric bright points. Plotted in the same format as Fig.\,\ref{fig:euitrc}a, but showing the FSI304 map obtained at 2025 March 21 UT\,21:09. The green contours within the yellow wedge region outline the bright points at that instance. Only those bright points that lived for at least 10\,hours are highlighted.\label{fig:euibp}}
 \end{center}
\end{figure}

Chromospheric and transition region network bright points are proxies of the magnetic field, which are advected by surface flows. These features can be used to track the flow field of the photosphere \citep[][]{2009A&A...495..319I}. To derive the long-term flow that is persistent for several hours (much longer compared to the typical granular lifetimes of 6\,minutes), we pursued the following approach. We first defined a wedge that subtended 120\textdegree\ angle at the zenith, extending to $\sim$30\textdegree\ from the south pole (yellow colored wedge in Fig.\,\ref{fig:euitrc}a). The mean and standard deviation of the FSI304 intensity as a function of time within this wedge are computed. Then in every snapshot, we retained only those pixels enclosed by the wedge with EUV intensity greater than 3$\sigma$ above the mean. This intensity threshold ensures that we are capturing only strong intensity features that are typically associated with network bright points at the coronal base. To fill any small holes in the detected features, we applied a morphological closing operator with kernel size of 3 in both space and time. We then labelled all the individual features and only selected those that have lifetimes of at least 10\,hours (90 in total detected over the course of eight days of EUI observations). Examples of these bright points are shown in Fig.\,\ref{fig:euibp}.

The intensity-weighted spatial position of each of these bright points is then calculated as a function of time. Because only the distances measured along the meridians (lines that pass through the zenith) are accurate in this projection, we first calculated the radial distance of each bright point at every timestep of its lifetime. This essentially yields their latitudinal position at each timestep. These latitudinal positions of all the 90 bright points as a function of time are identified with gray symbols in Fig.\,\ref{fig:euitrc}b. A linear fit is applied to these tracks. The slope of this line (in units of m\,s$^{-1}$) represents the latitudinal velocity of the bright point motion.

To further understand the photospheric flows as a function of latitude, we applied a Fourier local correlation tracking technique \citep[FLCT;][]{2004ApJ...610.1148W,2008ASPC..383..373F,2020ApJS..248....2F}. To this end, we first normalized the EUI data at each snapshot with the mean calculated from the wedge region (as described above). Then we remapped these {\sc arc} projected data to a polar ($r\phi$) projection with $r$ representing the latitude and $\phi$ the longitude. The grid scale along the radial direction is 0.1\textdegree\ consistent with the {\sc arc} projection, while in the longitudinal direction it is $\sim$0.2\textdegree. 

We used 7\,pixels as the Gaussian kernel size for the convolution operation and enabled the bias correction flag, both of which are inputs to the FLCT algorithm. The 7\,pixel window will smoothen out small structures covering only a few pixels, but will retain the supergranular structure that we are interested in. Only those features whose intensity is at least 0.65 in the mean-normalized images are tracked. This thresholding mainly discards the very dark regions in the images but retains network elements of generally varying brightness in space and time. This basically ensures that we are tracking only strong intensity features related to the magnetic network. The resulting latitudinal velocity maps are temporally and longitudinally averaged to compute the mean latitudinal flow at high latitudes. The longitudinal average was limited to the meridians of the wedge region in Fig.\,\ref{fig:euitrc}a. 

Moreover, because of the low number statistics of chromospheric network patches available for tracking very close to the pole, the flow signal will be noisy. The computed flow signal exactly at the pole is not meaningful because a single pixel on the EUI detector is turned into many pixels, each having a different longitude.   

Visual inspection of the mean velocity profile revealed three spurious isolated spikes in the radial profile, with velocity in excess of $-190$\,m\,s$^{-1}$ (i.e, directed toward the south pole). We simply replaced these by average values of the velocity derived from the respective adjacent pixels on both sides of those spikes. This artifact-corrected mean velocity as a function of solar latitude is shown in Fig.\,\ref{fig:euitrc}c. The latitudinal average value of this flow up to $-85$\textdegree, i.e., excluding the innermost $5$\textdegree\ latitude range from the pole,  is $-22$\,m\,s$^{-1}$. This implies a persistent flow toward the pole.

\end{document}